\newcommand{\beq}{\begin{quote}}
\newcommand{\enq}{\end{quote}}
\newcommand{\be}{\begin{equation}}
\newcommand{\en}{\end{equation}}

\documentstyle[12pt]{article}
\begin{document}
\title{Violation of an
inequality in an experimental test of Leggett's non-local
hidden variable theory  } 

\date{}
\author{}

\maketitle
\begin{abstract}
I impose   previously avoided symmetry conditions 
in Leggett's non-local hidden variable theory that are 
required by the conditions in
a recent experimental  test of this 
theory \cite{zeilinger}
These conditions  lead
to an inequality for the  polarization
correlation function that has a maximal violation 
from the quantum predictions that is  
2.5 times larger than reported in this article.
Furthermore, when these symmetry conditions are applicable,  
Leggett's  non-local theory  also cannot model the observed 
quantum  correlations in previous  experiments.                                                                          
\end{abstract}

\subsection*{}

In a recent article entitled {\it An experimental test of 
non-local realism} \cite{zeilinger},
Gr\H{o}blacher et al.  derived an inequality, 
based on Leggett's non-local hidden variable theory \cite{leggett}, 
for the mean value $E$
of the polarization correlation of two photons, and compare it with
their data.
For a {\it general} source producing mixtures of two photons,
this mean value is determined in Leggett's 
theory by a   distribution function 
$F(\bf u,\bf v) $, where  
$\bf u$ and $\bf v$ are the photon  polarization
unit vectors on the Poincar\`{e} sphere.
But according to quantum mechanics, 
in the reported  experiment \cite{zeilinger} 
an entangled two photon  state $\psi_{A,B}$  is created with zero angular momentum and  
{\it odd} parity,  which therefore has  the form  (see Appendix) \cite{zeilinger2}
\be
\label{odd1}
\psi_{A,B}=(1/\sqrt 2)[\psi_{A}(\bf u)\psi_{B}(-\bf u) -\psi_{A}(-\bf u)\psi_{B} (\bf u)].
\en
Hence,  these two photons always have
polarization vectors  directed in  opposite directions
on the Poincar\`{e} sphere, i.e.  $\bf v = -\bf u$. Furthermore,
this  state has the remarkable
property  that it is {\it independent} of the orientation
of $\bf u$ on this sphere (see Appendix). To maintain  both of these   properties 
in Leggett's theory requires that the 
distribution function 
\be
\label{fuv}
F(\bf u,\bf v)=\frac{1}{4\pi} \delta (\bf u + \bf v)
\en
 where $\delta$ is the Dirac delta function. Then, the hidden variable 
polarization correlation function  $E$  depends only on the
{\it relative}  orientation of the two  polarization analyzers
represented by unit vectors $\bf a$
and $\bf b$ on the Poincar\`{e} sphere. Substituting Eq. \ref{fuv}
in Leggett's inequality, equation (4.2) in reference  \cite {leggett} 
(with an incorrect  factor 2 deleted) gives
\be
\label{integral1}
-1+\frac{1}{4\pi}\int  d\bf u |\bf u \cdot (\bf a+ \bf b)|
\leq P(\bf a,\bf b) \leq 1 -\frac{1}{4\pi} \int d \bf u |\bf u \cdot (\bf a -\bf b)|.
\en
The angular integrations on both sides of this inequality
can be easily carried out by chosing
for the $z$ axis the directions of the vectors $\bf a +\bf b$ and $\bf a - \bf b$ respectively,
and one obtains

\be
 -1+|cos(\phi/2)| \leq -E(\phi) \leq 1-|sin(\phi/2)|,
\en
where $\phi$ is the angle between the unit vectors $\bf a$ and $\bf b$,
and $E(\phi)=-P((\bf a, \bf b)$.
This inequality is  more restrictive than the one derived
by Gr\H{o}blacher et al \cite{zeilinger}.

In quantum mechanics, one finds that $E$ has the form (see Appendix)
\be
E_Q(\phi)=-\bf a \cdot \bf b=-cos(\phi),
\en
which is in good agreement with the reported experimental results, but 
it violates this inequality  for $\phi$ in the ranges $0^o<\phi<60^o$ and $120^o<\phi<180^o$.
The maximal violation  occurs at $\phi=28.8^o$ 
and $\phi=151.2^o$, and its magnitude  is $2.5$ 
larger than the value obtained  by the inequality of  Gr\H{o}blacher et al. 
\cite{zeilinger}. Our inequality also 
contradicts  their claim that ``existing data of all Bell tests cannot
be used to test the class of non-local theories consider here''.

When the  entangled two photon state
of zero angular momentum has {\it even} parity, 
as is the case, for example, with the photon pairs created in the 
first experiments testing Bell's inequality \cite{clauser},\cite{aspect}, I find that (see Appendix)
\be
\label{correven}
E_Q=\bf a \cdot \bf b -2\bf a_z \bf b_z, 
\en 
where the z-axis in the Poincar\`{e} sphere is chosen to represent circular polarization.
In the original experiments, the analyzers  measured only  linear
polarization for which  $\bf a_z=\bf b_z=0$.
It would be of interest in this case  also to  verify experimentally 
the correlations for elliptical polarization  when 
$\bf a_z$ and $\bf b_z$ do not vanish.  

\subsection*{Appendix. Two photon polarization correlation function in quantum mechanics}

Let $\psi^{+}$ represent the state of right handed circular polarization 
along the $z-$ axis of the Poincar\`{e}
sphere, and $\psi^{-}$ the corresponding state of left handed circular polarization. Then
\be
\psi^{\pm}=\frac{1}{\sqrt{2}}(\bf e_x \pm i \bf e_y)
\en
where $\bf e_x$ is a unit vector for linear polarization along the $x$ axis,
and $\bf e_y$ is the corresponding vector for the $y$ axis.
A general state of elliptic polarization $\psi(\bf u)$ is given by
\be
\label{rep}
\psi(\bf u)=e^{-i\phi/2}cos(\theta/2)\psi^{+}+e^{i\phi /2}sin(\theta/2)\psi^{-},
\en
where $\theta, \phi$ are the polar angles representing the unit vector $\bf u $ on the Poincar\`{e}
sphere.  This  state is an eigenstate of the Pauli spin matrix $\sigma \cdot \bf u$ along the
unit vector $\bf u$ with eigenvalue  1,
\be
\bf {\sigma} \cdot \bf u \, \psi(\bf u)=\psi(u), 
\en
and it  has the useful property 
\be
\label{cute}
(\psi(\bf a)|\sigma \cdot \bf b |\psi (\bf a))= \bf a \cdot \bf b .
\en

The  polarization correlation function $E_{A,B}$ for a general two photon state $\psi_{A,B}(\bf u,\bf v)$
is given by
\be
\label{corrfunction}
E_{A,B}=(\psi_{A,B} |\sigma_{A} \cdot \bf a \,\sigma _{B} \cdot {\bf b}| \psi _{A,B}),
\en
where $\bf a, \bf b$ are the elliptical polarization analyzers for photons $A,B$. 
 The parity transformation $P$ is defined by
\be
P \psi(\bf u)=-i \psi(-\bf u)
\en
and $P^2=1$.
Hence, the state $\psi_{A,B}$ defined by Eq. \ref {odd1} is a state of odd parity, and
substituting the representation Eq. \ref{rep} for $\psi(\bf u)$ into this equation one obtains
\be
\label{oddlabel}
\psi^{odd}_{A,B}=-\frac{i}{\sqrt(2)} (\psi^{+}_{A}\psi^{-}_{B} -\psi^{-}_{A} \psi^{+}_{B})
\en
which demonstrates that this state is independent of the direction of the unit vector  $\bf u$.
Substituting this state in Eq. \ref {corrfunction} one obtains
\be
E^{odd}_{A,B}=-\bf a \cdot \bf b.
\en
The corresponding two photon state of even parity is
\be
\psi^{even}_{A,B}=-\frac{i}{\sqrt(2)} (\psi^{+}_{A}\psi^{-}_{B} +\psi^{-}_{A} \psi^{+}_{B}),
\en
and substituting in Eq. \ref {corrfunction} one  obtains 
\be
E^{even}_{A,B}=\bf a \cdot \bf b -2 \bf a_z \bf b_z.
\en

\subsection*{Acknowledgment}
I would like to thank one of the reviewers of an earlier version of this manuscript for
suggesting several improvements.


\begin{thebibliography}{30}
\bibitem{zeilinger2}   In reference \cite{zeilinger},  under the caption of 
                    Figure 2,  the authors report that in their experiment `` the crystal 
                    is aligned to produce the polarization
                    entangled singlet state''. This state is written  in a basis for circularly
                    polarized states aligned along the z-axis of their  Poincar\`{e} sphere,  which
                    corresponds to  Eq. \ref{oddlabel} in the Appendix of this paper .  This  particular
                    representation, however, does not reveal the symmetry property of this state
                    which is shown in this Appendix to be independent of the choice of  polarization
                    of the basis state.
\end{thebibliography}
\end{document}